# Supernovae and the Chirality of the Amino Acids

R.N. Boyd[1]*, T. Kajino[2], and T. Onaka[3]

[1]Lawrence Livermore National Laboratory
[2]National Astronomical Observatory of Japan, and University of Tokyo, Department of Astronomy, Graduate School of Science
[3]University of Tokyo, Department of Astronomy, Graduate School of Science

*To whom correspondence should be directed:
Richard N. Boyd
Lawrence Livermore National Laboratory
PO Box 808, L-414
 Livermore, CA 94551
925-423-3201
FAX: 925-292-5940
 boyd11@llnl.gov

Running Title: Amino Acid Chirality

Abstract

A mechanism for creating amino acid enantiomerism that always selects the same global chirality is identified, and subsequent chemical replication and galactic mixing that would populate the galaxy with the predominant species is described. This involves: (1) the spin of the $^{14}$N in the amino acids, or in precursor molecules from which amino acids might be formed, coupling to the chirality of the molecules; 2) the neutrinos emitted from the supernova, together with magnetic field from the nascent neutron star or black hole formed from the supernova selectively destroying one orientation of the $^{14}$N, and thus selecting the chirality associated with the other $^{14}$N orientation; (3) chemical evolution, by which the molecules replicate and evolve to more complex forms of a single chirality on a relatively short timescale; and (4) galactic mixing on a longer timescale mixing the selected molecules throughout the galaxy.

1. Introduction

A longstanding puzzle in astrobiology, and in other fields with which astrobiology interfaces, has been the existence of left-handed amino acids and the virtual exclusion of their right-handed forms. This is especially puzzling because most mechanisms suggested for creating this "enantiomerism" would create one form in nature locally but would create equal numbers of the other somewhere else. The total dominance of the left-handed forms on Earth is well known, but the left-handed forms appear also, based on limited statistics, to be preferred in the cosmos. While left-handed excesses of some amino acids were reported in the Murchison meteorite *(Kvenvolden et al. 1970; Bada et al. 1983; Cronin et al. 1988; Cronin and Pizzarello 1997)*, not all detected amino acids were found to be enantioimeric. This latter observation is especially interesting for two reasons: the preference of the left-handed forms apparently persists throughout the cosmos, but both forms were apparently frozen into the meteorite before the right-handed forms could be eliminated. Recent work *(Chyba 1990)* has confirmed the preference for the left-handed forms.

It is generally accepted that if some mechanism can introduce an imbalance in the populations of the left- and right-handed forms of any amino acid *( Glavin and Dworkin 2009)*, successive synthesis or evolution of the molecules will amplify this enantiomerism to produce ultimately a single form. Indeed, the initial imbalance is as small as a part in $10^{17}$ in some models *(Bonner 1996)*. What is not well understood, though, is the mechanism by which the initial imbalance can be produced, and the means by which it always produces the same chirality, i.e., the left-handed form for the amino acids. The energy states of the left- and right-handed forms have been shown, by detailed computations, to differ at most by infinitesimal amounts *(see, e.g., Tranter and MacDermott 1986; Klussmann et al. 2006)*, so it would be difficult for thermal equilibrium to produce the imbalance. However, it also appears that transitions between the two forms are difficult (at least in gas or solid phases), so despite their tiny energy differences some much larger barrier apparently must be surmounted for transitions to occur once an imbalance is established.

One suggested mechanism lies with processing of a population of amino acids, or of their chiral precursors, by circularly polarized light *(see, e.g., Bailey et al.*

*1998; Takano et al. 2007)*; this could select one chirality over the other. However, this solution does not easily explain why it would select the same chirality in every situation, or why the physical conditions that would select one form in one place would not select the other in a different location. Another possibility *(Mann and Primakoff 1981)* invoked selective processing by some manifestation of the weak interaction, which does violate parity conservation, so might perform a selective processing effect that could ultimately be related to molecular chirality. This idea was based on earlier work by *Vester et al. (1959)*, and *Ulbricht and Vester (1962). Mann and Primakoff (1981)* focused on the β-decay of $^{14}$C to produce the selective processing. However, it was not possible in that study to show how simple β-decay could produce chiral-selective molecular destruction. Another suggestion *(Cline 2005)* involving the weak interaction assumed that neutrinos emitted by a core-collapse supernova would selectively process the carbon or the hydrogen in the amino acids to produce enantiomerism. Although this suggestion benefits from the naturalness of the weak interaction in producing chirality and of supernovae to process large regions of space, it also has difficulty explaining how any predisposition toward one or the other molecular chirality could evolve from the neutrino interactions. A similar suggestion *(Bargueno and Perez de Tudela 2007)* involves the effects of neutrinos from supernovae on electrons.

In this paper we also invoke the weak interaction to perform selective destruction of one chirality or symmetry of molecules. The key to this mechanism would be the selective processing of the molecules that have been observed to exist *(see, e.g., Knacke et al. 1982)* in molecular clouds in the vicinity of a core-collapse supernova, but would be specifically related to $^{14}$N, a nucleus with spin of 1 (in units of Planck's constant, $\hbar$) that is common to all amino acids. Two features of supernovae are important; one being the magnetic field that would be established as the star collapsed to its final state, a neutron star or a black hole, and the other being the intense flux of neutrinos that would be emitted as the star collapsed. Although neutron stars and black holes are known to generate extraordinarily strong magnetic fields, the fields from these stars would presumably not have reached their full strength within the few seconds during which neutrinos are emitted. However, the fields in question would only have to couple to the molecules strongly enough to produce some orientation of their non-spin-zero nuclei. This will be discussed in Section 2.

The $^{14}$N must couple in some way to the chirality of the molecules; we assume a model to describe this. Although this model is not unique, we will assume that it is representative of the effects it produces, and that the $^{14}$N spin is correlated with the molecular chirality in a way that is consistent through all the amino acids. Then, the neutrinos will preferentially interact with the $^{14}$N atoms in one of the chiral forms, and convert the $^{14}$N to $^{14}$C, thereby destroying that molecule and thus preferentially selecting the other chiral form. This will be the subject of Section 3.

Another question involves the extent to which a single chirality might populate the entire Galaxy. We show that it would be unlikely for supernovae to do so by themselves, but that subsequent chemical amplification of the chirality-selected, biologically-interesting molecules would drive the dominant form toward total local dominance. Then Galactic mixing, operating on a slower timescale, would be able to establish the dominant form throughout the Galaxy. These two mechanisms would make it likely that the Galaxy would be populated everywhere with a single form, or at least that the same form would dominate at all Galactic locations. These issues are described in Section 4.

In Section 5 we address two issues that could arise from considerations of this model, one being the possibility that other nuclei in the amino acids might produce similar effects to those put forth for $^{14}$N, and the other that the photon flux from the supernovae that are producing the chiral selection, or from their progenitor stars, might also destroy all the molecules that were selected.

Section 6 discusses experiments that might be performed to test some of the assertions of this model. Finally, Section 7 presents our conclusions.

The various mechanisms by which this model functions are illustrated in figure 1.

2. Selective Destruction Mechanism

The key nucleus in the selective processing of this model is $^{14}$N. Although it is the spatial arrangement of the atoms that determine the chirality of the molecule, the $^{14}$N nuclear spin would couple to the electronic spin, which would couple to an external magnetic field, thereby aligning the spin 1 nuclei also to that magnetic field throughout a large volume of space, and providing an orientation direction of the $^{14}$N spin in the molecular environment. The separation between magnetic substates μB, where μ is the magnetic moment of $^{14}$N, of order 6x10$^{-7}$ eV/G, and B is the magnetic field generated by the residual electron spin. This would generate a hyperfine interaction that would split the magnetic substates of the nucleus+electron spin. Indeed, as discussed in the next section, this splitting could be sizable if the electrons exhibit collective effects. The thermal energy kT for galactic molecular clouds, generally assumed to be at around 10-20K *(Ferriere 2001))*, would be about 10$^{-3}$ eV. Thus some orientation of the $^{14}$N would be expected to occur from the interaction with the neutron star's or black hole's magnetic field, even at fairly large distances from the center of the star.

The other important feature of this scenario is that of the chirality of the neutrinos: electron antineutrinos, $\underline{\nu}_e$s, are right-handed and electron neutrinos, $\nu_e$s, are left-handed (for discussions of neutrino properties see *Frauenfelder and Henley 1974, Boyd 2008*). Near the neutron star or black hole the spin of the $^{14}$N would be aligned with that of the electron antineutrino, so the total neutrino-nuclear spin would have to be 3/2, but at the opposite side of the star the antineutrino and $^{14}$N spins would be antialigned, producing a mixed total spin, and a different magnetic substate distribution. Then the reaction

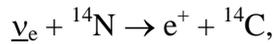

$\underline{\nu}_e + {}^{14}N \rightarrow e^+ + {}^{14}C,$

would not have the same strength on opposite ends of the neutron star or black hole, as is discussed below, and the resulting destruction of the $^{14}$N, and hence of the molecule, would depend on the electronic structure, hence the chirality or symmetry, of the molecule. The above transition is pure Gamow-Teller *(see, e.g., Boyd 2008)*, since it is between nuclei with spins that differ by one unit (this also applies to the $\nu_e + {}^{14}N \rightarrow e^- + {}^{14}O$ reaction, discussed below). This situation is illustrated in figure 2.

The Q-values of the reactions are critical to this argument; that for $\underline{\nu}_e + {}^{14}N \rightarrow e^+ + {}^{14}C$ is -1.18 MeV, and for $\nu_e + {}^{14}N \rightarrow e^- + {}^{14}O$ is -5.14 MeV. The high threshold energy for the latter reaction would inhibit it significantly compared to the former, as the energies of the neutrinos emitted in the collapse to a neutron star are comparable to that Q-value *(Keil et al. 2003; Esteban-Pretel et al. 2008; Duan et al. 2006; Duan et al. 2007; Duan et al. 2007a)*. In addition, the cross sections for the neutrino induced reactions increase as the square of the energy above threshold (*Boyd 2008*). Thus the $\nu_e$ induced reaction would not reverse the effect produced by the $\underline{\nu}_e$s, assuming the two species have comparable energy distributions. While many electron neutrinos would still be capable of converting $^{14}$N to $^{14}$O, many more electron antineutrinos would be capable of driving $^{14}$N to $^{14}$C producing a chirally preferred mode by preferentially destroying the molecules with the other chirality. Although the neutrinos emitted in collapse to a black hole may be somewhat higher in energy *(Lunardini 2009)* than those from a neutron star, they are not sufficiently dissimilar to change this conclusion.

The magnitude of the selection in the extreme situations for $\underline{\nu}_e + {}^{14}\text{N} \to e^+ + {}^{14}\text{C}$, i.e., those in which the antineutrino spin and magnetic field are either aligned or antialigned, can be estimated as follows. Assuming that the spins and magnetic field are aligned, then the angular momentum of the combined $^{14}$N(spin 1) + $\underline{\nu}_e$(1/2) system must be 3/2, whereas if they are antialigned then it will be ½. The transition produces $^{14}$C(0). Thus, since the weak interaction can be assumed to be point-like for the present situation, conservation of angular momentum in the aligned case requires some orbital angular momentum transfer. The weak interaction does violate parity, so this can be $\ell=1$ (again, in units of Planck's constant, $\hbar$), but this will inhibit the transition compared to that for the antialigned case, in which angular momentum can be conserved with no orbital angular momentum transfer, by roughly an order of magnitude. Since the angular momentum coupling coefficients favor the case in which the spins of the $^{14}$N and $\underline{\nu}_e$ are antialigned by 2:1 over that in which they are aligned, the antialigned case $^{14}$N nuclei will suffer destruction roughly a factor of seven (two-thirds of the order of magnitude) greater than that in which they are aligned.

Determination of the total effect on the selective destruction of the chiral molecules would require integration over all space and a detailed accounting of the magnetic fields. However, the average effect may not be relevant. If amplification of the enantiomerism produced by any effect has a threshold value, or if it progresses more rapidly where the enantiomeric excess is larger, the relevant number will most likely be the maximum enantiomeric excess achieved. We return to this point at the end of section 5. The effect discussed would inevitably produce significant enantiomerism of the chiral molecules and integration over all space is performed, this mechanism would select the same overall chirality for destruction *for every core-collapse supernova*. Note that the excited states of $^{14}$C and $^{14}$O would not be expected to play much of a role in the destruction of $^{14}$N; the first excited states of both nuclei are at about 6 MeV, so the Q-values for those transitions would be much more negative than those to the ground states.

3. Coupling of the nuclear spins to molecular chirality

To illustrate how the $^{14}$N spin might couple to the chirality of the molecules in which they are contained, we have considered several models that might apply (see a general discussion in *Avalos et al. 1998* and, e.g., *Nafie and Freedman 1986*). The one that seems most relevant for our situation was developed in the context of nuclear magnetic resonance (*Buckingham 2004; Buckingham and Fischer 2006*). In this model, as shown by *Faraday (1846)*, the magnetic field will rotate the plane of polarization of light traveling through molecules (this applies to both chiral and achiral matter). Paraphrasing from *Buckingham (2004)*, the magnetic field induces a current density which, in chiral samples, causes a nuclear magnetic moment (odd under time reversal and even parity) to induce a molecular electric dipole moment (even under time reversal and odd parity). The result is an effect that is opposite in sign for left-handed and right-handed enantiomers.

The critical feature for this model (*Buckingham 2004*) is the necessity of having a non-zero spin nucleus, i.e., the $^{14}$N, for this effect to occur at all. While it might be possible to couple that $^{14}$N spin to the molecular chirality in other models, it is clearly essential in this model.

In order to assume greater specificity, our discussion would now have to look at specific amino acids, or at their precursor molecules, in detail. This would be well beyond the scope of this paper, so we will assume that the above general statements apply to all amino acids (with the possible exception of glycine, which is non-chiral), and discuss the general effects which might be expected. This may be a better assumption for amino acids with a single chiral center than for those with an additional center.

4. Spreading Enantiomerism Throughout the Galaxy

To what extent would all the molecules in the Galaxy be processed by the effect being described? The supernovae by themselves cannot perform this task. The molecules are thought to be created and contained in the molecular clouds that pervade the Galaxy, and the bulk of the core-collapse supernovae also occur there. The frequency of core-collapse supernovae is roughly one every 30 years in the Galaxy *(Beacom et al. 2001)*, or $3\times10^5$ during the lifetime of a molecular cloud, roughly *(Elmegreen 2000; Hartmann et al. 2001)* 10 My. The magnetic dipole field falls off as $1/r^3$, and neutron star magnetic fields are thought to be as high as $10^{14}$ gauss at their surface *(Duncan and Thompson 1992)* of radius ~10 km. The field that orients the molecules must be greater than the ambient interstellar magnetic field *(Ferriere 2001)*, typically around $10^{-6}$ gauss Thus the cloud's molecules would be oriented if they were within ~$10^{13}$ cm of the center of the supernova, and the volume processed by the $3\times10^5$ supernovae, assuming their optimal spacing, would be $1\times10^{45}$ cm$^3$. Roughly $2\times10^9$ solar masses of material is in the Galactic molecular clouds *(Ferriere 2001)*, at a density ranging from $10^2$ to $10^6$ cm$^{-3}$, giving (assuming $10^4$ cm$^{-3}$) a total cloud volume of $2.4\times10^{62}$ cm$^3$, a value far greater than that which could be processed by the supernovae that would occur during the cloud's lifetime.

The supernova neutrino flux would be sufficient to process the $^{14}$N nuclei at the largest distances from the supernovae at which the magnetic field would provide orientation. At a distance of $10^{12}$ cm from the supernova the electron antineutrino flux will be roughly $10^{31}$ cm$^{-2}$. Estimating a cross section for the destruction reaction on $^{14}$N of $10^{-40}$ cm$^2$ (*Fuller, Haxton and McLaughlin 1999*) gives a probability of $10^{-9}$ for each nucleus to be processed. Closer distances would provide an even higher probability of processing, but this is limited by the proximity to the progenitor star as is discussed further below. However, this is still a very small enantiomeric excess. Thus we believe that a more plausible mechanism for driving the entire Galaxy to some level of enantiomerism is a combination of chemical evolution and Galactic mixing. Indeed we believe that these mechanisms are unavoidable. As soon as a supernova explodes the enantiomeric material it has produced begins to mix with the racemic (populated equally with both chiralities) material adjacent to its processed volume, driving the racemic material with which it mixes toward partial enantiomerism, possibly on the surfaces of the dust grains *(see, e.g., Hasegawa et al. 1992)* that also exist in the molecular clouds. This ultimately would extend the enantiomeric volume well beyond that which was processed by the supernova neutrinos.

The ability of a collection of molecules exhibiting a small enantiomeric excess to amplify that excess dramatically has been demonstrated, at least in some environments. A general discussion of this capability has been given by *Kondepudi and Nelson (1985)*, who showed theoretically how such enantiomeric excess might be produced by weak neutral currents, even from fluctuations, as discussed by *Mason and Tranter (1984),* and then amplified toward much greater enantioimerism by chemical replication. The chemical replication of molecule X was suggested by *Mason (1984)* and *Kondeputi and Nelson (1985)* to proceed by autocatalysis, and to preserve chirality, for example, by S+T↔X, which then has its chirality established so to become $X_L$. Replication and amplification then could occur by S+T+$X_L$↔2$X_L$. The ability of autocatalysis to produce amplification of enatiomerism was demonstrated in the laboratory by *Soai et al. (1995) and Soai and Sato (2002).* A subsequent study by *Mathew, Iwamura, and Blackmond (2004)* confirmed the possibility of autocatalysis, and did so in an environment that was more relevant to amino acid autocatalysis. *Breslow and Levine (2009)* showed that amino acid enantiomerism could be amplified by successive evaporations to precipitate the racemate, with the solution becoming highly enantiomeric. Although this latter mechanism would probably not be relevant to amplification in dust grains, it could well produce amplification once the somewhat enantiomeric grains or meteorites landed in a suitable planetary environment.

Perhaps more specific to cosmic amplification, *Garrod et al. (2008)* have developed a model in which chiral replication of complex molecules would occur in the interstellar medium in the warmed (possibly by the neutrinos, albeit for a short amount of time) ice outer shells of grains. In their model, chemical replication would be catalyzed by radicals, for example, H, OH, CO, $CH_3$, NH, and $NH_2$, created by the interactions of high energy cosmic rays with preexisting molecules. Subsequent studies might involve radicals that are more complex, and more relevant to amino acid replication. The grains would have to be warmed over the ambient temperature of the molecular clouds, perhaps to 200 K, so as to enhance the mobility of the heavy radicals.

We believe that either autocatalysis or chemical replication via radicals could produce the chemical replication required for the present model to succeed, and might well be expected to maintain the chirality established by the magnetic fields and neutrinos from supernovae.

As the enantiomeric excess of the processed material increases via chemical replication, it will also mix with the rest of the material in the Galaxy, ultimately establishing at least a preference for left-handed amino acids throughout the Galaxy. Details of the processes by which this would occur have been discussed by *Pittard (2007)* in much greater detail than is appropriate for the present paper, and include "many types of astronomical sources, including planetary nebulae, wind-blown bubbles, supernova remnants, starburst superwinds, and the intracluster medium". Suffice it to say that significant mixing would occur. This mixing would be expected to occur on a much slower timescale than the chemical processing timescale, as discussed below, but would inevitably establish a preferred chirality throughout the Galaxy.

It is important to consider the timescales of the chemical evolution and the galactic mixing for the present model to produce at least some enantiomerism throughout the galaxy on a timescale less than the age of the universe. Although abundance differences surely do exist within the galaxy, it is generally assumed that the galactic mixing timescale is much smaller than the age of the universe *(Bennett et al. 2003)*. As one signature of galactic mixing time, our Galaxy rotates roughly once every $3 \times 10^8$ years *(see, e.g., Kulkarni et al. 1982)*, much less than the $\sim 12 \times 10^9$ years the Galaxy has lived. The evolutionary timescales of organic molecules are undoubtedly much shorter than the galactic mixing timescales. Although this might depend on many variables, the fact that such molecules are born in the molecular clouds, and that these clouds are born, live, and die in of order 10 My *(see, e.g., Elmegreen 2000; Hartmann et al. 2001)*, confirms the shortness of the chemical evolutionary timescale.

While it might be thought unnecessary to speculate about the enantiomerism that this model would spread throughout the Galaxy, based on the tiny number of meteoritic samples, and Earthly living beings, that we have, the model we discuss above for propagating the supernova-selected enantiomerism throughout the Galaxy allows for the same chirality to exist on a large scale. However, as noted in section 2, the electron neutrinos will convert some $^{14}$N to $^{14}$O, which would presumably select right handed chiral molecules. The numbers of these right handed molecules would be appreciably less than those of the left handed ones, but they might be observed in, for example, the meteorites in which enantiomerism, but not homochirality, is found to exist. One new test should occur before long; it is hoped that the Japanese satellite Hayabusa *(Fujiwara et al. 2006; Yano et al. 2006; Saito et al. 2006)* will return to Earth in 2010, and may have some dust samples from asteroid Itokawa; all amino acids included in those samples must also have chiralities consistent with those of their terrestrial counterparts. When these samples have been analyzed for their amino acids, this information should provide a strong test of the present model. An additional test will be provided in 2014 when the ROSETTA mission sends a lander onto Comet 67 P/Churyumov-Gerasimenko *(Thiemann and Meierhenrich 2001)*.

5. Potential issues for this model

Several questions immediately arise. One involves $^2$H, another spin 1 nucleus, and a $1.5 \times 10^{-4}$ component of hydrogen. However, the Q-values for creating 2 neutrons, from a $\underline{\nu}_e$ induced reaction, or 2 protons from a $\nu_e$ induced reaction, are not sufficiently different to strongly favor one reaction, so that the neutrinos on one side of the neutron star would have about the same effect as the antineutrinos on the other side. In addition, the $^2$H nuclei would be considerably less prevalent in the chiral molecules than would $^{14}$N, so they would not be expected to produce the same effect as would $^{14}$N. One might also ask if $^1$H could have a similar effect. However, the extremes of coupling of the neutrino spin to that of the proton would be much less significant in selectively destroying spin ½ nuclei than they would of spin 1 nuclei. Since the spin of the proton is ½, it would couple with the neutrino to form states of either total spin 1 or 0, the transitions would be a mixture of Gamow-Teller and Fermi *(Boyd 2008)*, and a strong selection effect would not be anticipated.

Another possible concern with this model is with the extraordinary photon field from the supernova to which the molecules, and the grains on which they are thought to form *(Jenkins and Savage 1974; Bohlin 1975)*, would be subjected. This would begin several hours after the neutrinos performed their processing, and might well destroy all the molecules that had been processed in the ten seconds during which the neutrinos were emitted. Some molecules might survive, as density fluctuations in some regions might provide the necessary attenuation of the photon flux. More significantly, however, some supernova will result in collapse to a black hole. (Note that we are discussing "ordinary" supernovae here, not those, presumably rarer, ones that might have an accretion disk, and/or might produce a gamma ray burst.) This has been studied recently *(Lunardini 2009; Heger et al. 2003)*; that study summarized previous work that suggests that stars having masses from 8 to 25 times that of the Sun would be expected to form a neutron star, stars from 25 to 40 solar masses would form a neutron star, but the fallback would ultimately produce a black hole, and stars having masses more than 40 solar masses would collapse directly to a black hole. *Lunardini (2009)* also showed that roughly the same number of neutrinos would be emitted from stars in the latter mass range as in the case in which the collapse goes to a neutron star, but that a larger fraction of the neutrinos would be electron antineutrinos and that their energies would be somewhat higher than in the supernovae from less massive progenitor stars. Both of these effects would enhance the chiral processing. In any event, for all these stars, the result would be a "failed supernova", i.e., the photon flux would be small or nonexistent. This would apply also to the $^{56}$Fe, which would power the supernova light curve were most of it not also swallowed by the black hole (Heger et al. 2003). The stars in the 25 to 40 solar mass range would produce some photons, but they would be suppressed compared to the neutron-star case. They might produce even more neutrinos than the stars that produce neutron stars *(Fryer 2009)*.

Apparently the supernovae that go directly to black holes would both produce the neutrinos required to produce molecular enantiomerism and not then destroy the molecules with photons. Because the time required for fallback black hole formation is short, many of these black hole forming supernovae *(Sumiyoshi et al. 2007)* might also be expected to produce sufficiently few photons that they would also not destroy the molecules that their neutrinos had processed. Supernovae that produce a neutron star are more frequent than those that produce a black hole *(Beacom et al. 2001)*, but the fraction that does produce a black hole is significant (~20% *[Lunardini 2009]*.

Another potential concern is with the progenitor star. A red giant would encompass the entire region that the supernova could process; these would be the progenitors of the 8-25 solar mass stars. Wolf-Rayet stars, thought to be the progenitors of the supernovae resulting from stars of more than 25 solar masses, are small, but are very hot. They have been observed to produce grains from their winds, but these would exist inside the clouds resulting from the winds, which are too hot to sustain molecular formation (*Crowther 2007*). However, dust grains or meteoroids that passed through the Wolf-Rayet clouds would be exposed to the highest temperature regions of the clouds through which they passed for much shorter periods of time. These would have been formed elsewhere, and could contain molecules formed as they passed through their giant molecular cloud. And there would be a large number of these grains and meteoroids that could produce enantiomerism. They only have to approach within ~one AU of the Wolf-Rayet star to be processed, and this is certainly a large enough volume to include a huge number of grains and meteoroids.

However, molecules on the grain or meteoroid surfaces would certainly get evaporated, and could be dissociated, once they reach the vicinity of the star (less than a few AU). But if they consisted of agglomerated grains, the molecules that resided on the surfaces of the internal grains could be retained in the grain. Such agglomerations would have to be large enough initially to withstand some radiative ablation as they passed through the Wolf-Rayet cloud, with the extent of the surface ablation depending on their closeness of approach to the central star. None the less, these grains and meteoroids appear to provide the best opportunity for relatively highly enantiomeric samples to be produced. An object passing within $10^{12}$ cm of the star would achieve a processing probability of roughly $10^{-9}$, and therefore, an enantiomeric excess of $0.5 \times 10^{-9}$. If it were large enough to survive if it passed by at $10^{11}$ cm, it would achieve a maximum enantiomeric excess of roughly $5 \times 10^{-6}$ percent, or possibly even higher, given the possible enhancements of electron antineutrinos from the massive supernovae.

6. Additional tests of this model

Although difficult, some experimental tests of some aspects of this model might be feasible. While it would be ideal to produce a racemic mixture of some amino acid, and expose it to an intense neutrino flux of appropriate energy (at, e.g., the Spallation Neutron Source) in the presence of an external magnetic field, then measure the ratio of right and left handed molecules after an appropriate time, this does not appear to be feasible. However, it might be feasible to expose an amino acid mixture to an intense neutrino flux, in the presence of varying magnetic fields, to attempt to observe the selectivity of production of the daughters of $^{14}$N, $^{14}$C and $^{14}$O, to see if their production rates depend on the orientation of the magnetic field and the neutrino direction. While this would not produce as convincing a confirmation of the above ideas as actual detection of enantiomerism, it would test the viability of the basic features of this model, and provide a way to provide an actual estimate of the extent to which enantiomerism could be produced by core collapse supernovae. None the less we emphasize that one or the other chirality *will* be selected by the interactions with the supernova neutrinos, and whichever one is chosen will ultimately pervade every region of every galaxy; there will be no galaxies somewhere where right-handed amino acids dominate.

7. Conclusions

If this model turns out to be correct, the longstanding question of how the organic molecules necessary to create and sustain life on Earth were created will have undergone a strong suggestion that the molecules of life were not created on Earth at all. Rather they would appear to have been created in the molecular clouds of the galaxy, with their enantiomerism determined by supernovae, and subsequently either transported to Earth in meteorites, swept up as the Earth passed through molecular clouds, or included in the mixture that formed Earth when the planets were created. Any scenario in which these molecules were created on Earth, e.g., that suggested by *Miller 1953*, would find it much more difficult to explain the enantiomerism that is observed on Earth and, apparently, generally in the cosmos.

Acknowledgements
This work has been supported by the US National Science Foundation grant PHY-9901241, and under the auspices of the Lawrence Livermore National Security, LLC, (LLNS) under Contract No. DE-AC52-07NA27344. This paper is LLNL-ABS-412217. The authors express their gratitude for helpful discussions with N. Sleep, I. Tanihata, R. Kuroda, T. Land, L.D. Barron, L. Fried, and E. Branscomb. They also very much appreciate the comments of the anonymous reviewers.

Author Disclosure Statement

No competing financial interests exist for any of the authors of this paper.

Figure captions

Figure 1. The various processes involved in producing enantiomeric molecules throughout the Galaxy.

Figure 2. Schematic diagram of the magnetic fields, $\underline{B}$, surrounding a nascent neutron star and the spins of the $^{14}$N nuclei, $S_N$, and of the neutrinos, $S_\upsilon$, emitted from the supernova that created the neutron star.

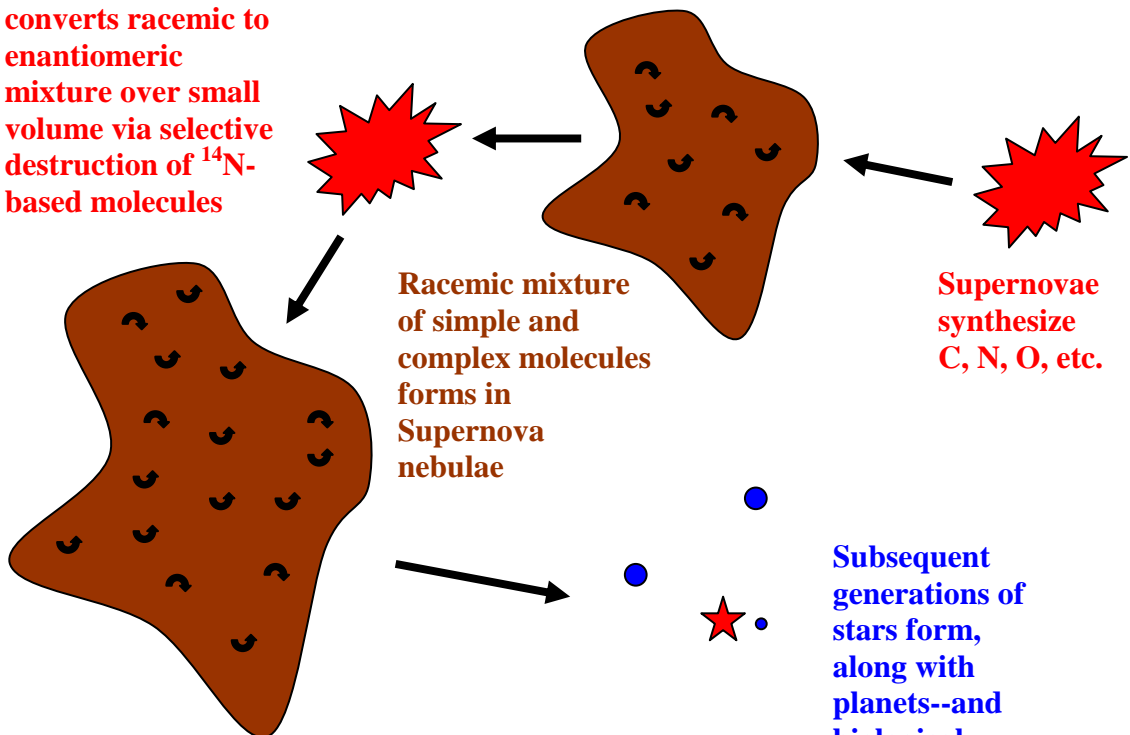

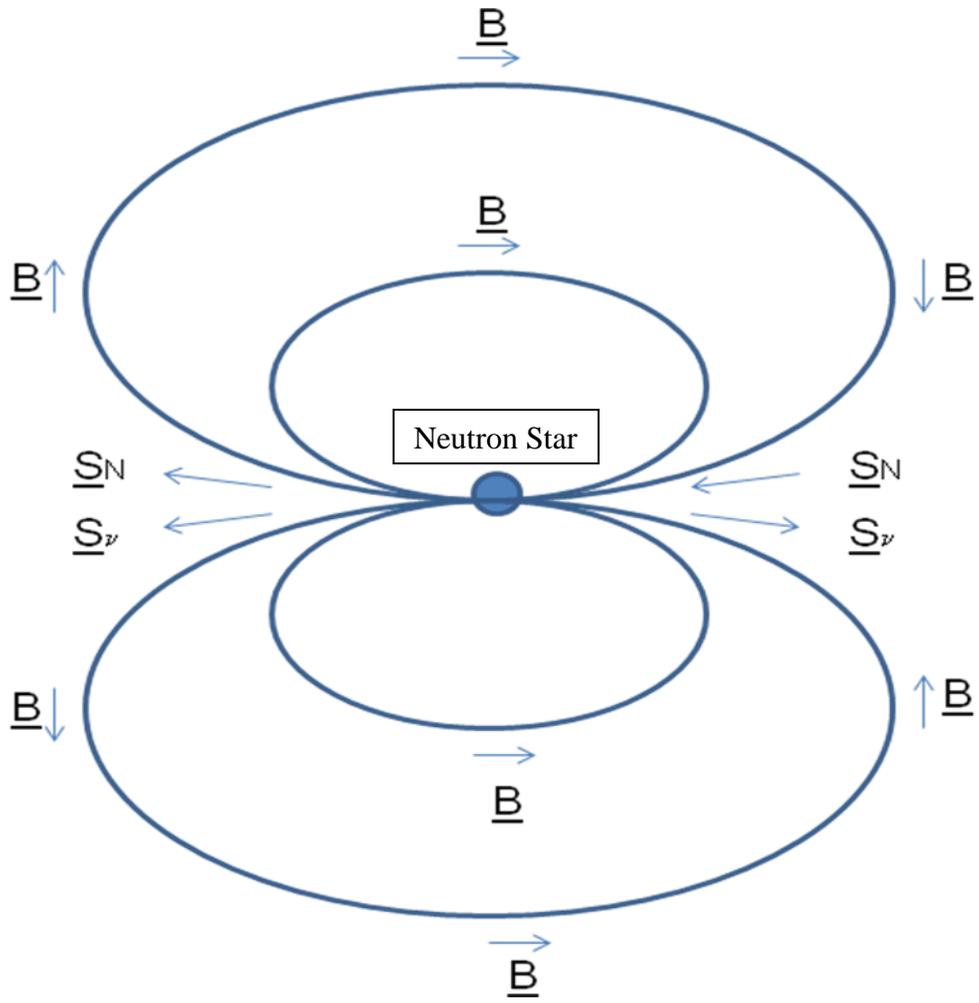